\newcommand{\avg}[1]{\langle{#1}\rangle}
\newcommand{\req}[1]{(\ref{#1})}

\newcommand{\beq}{\begin{equation}}
\newcommand{\eeq}{\end{equation}}
\newcommand{\beqar}{\begin{eqnarray}}
\newcommand{\eeqar}{\end{eqnarray}}

\documentstyle[aps,prl,twocolumn]{revtex}
\begin{document}
\title{Universal Behavior of One-Dimensional
Gapped Antiferromagnets in Staggered Magnetic
Field}
\author{Sergei Maslov and Andrey Zheludev}
\address{Department of Physics,
Brookhaven National Laboratory, Upton, NY 11973}
\date{\today}
\maketitle
\begin{abstract}
We study the properties of one-dimensional gapped Heisenberg
antiferromagnets in the presence of an arbitrary strong staggered
magnetic field. For these systems we predict a universal form for the
staggered magnetization curve. This function, as well as the effect the
staggered field has on the energy gaps in longitudinal and transversal 
excitation spectra, are determined from the universal form of the effective 
potential in $O(3)$-symmetric 1+1--dimensional field theory. 
Our theoretical findings
are in excellent agreement with recent neutron scattering data on
$R_2$BaNiO$_5$ ($R$ = magnetic rare earth) linear-chain mixed spin
antiferromagnets.
\end{abstract}
\narrowtext
One-dimensional isotropic Heisenberg antiferromagnets with an 
exchange gap in the
magnetic excitation spectrum have been at the center of theoretical and
experimental attention for almost two decades. This class of materials
includes {\it integer}-spin Heisenberg chains\cite{Haldane83} 
(commonly referred to as
Haldane-gap systems) and half-integer spin ladders with an {\it even}
number of legs\cite{Dagotto96}. Due to the presence of strong
quantum fluctuations in these systems the staggered magnetization has a
finite correlation length. The principal feature of the excitation
spectrum is a degenerate triplet of sharp spin-1 excitations commonly
referred to as magnons, separated from the ground state by a finite
energy gap $\Delta$. In recent years much work was aimed at
understanding the behavior of such gapped 1D antiferromagnets in the
presence of an external {\it uniform} magnetic
field\cite{Affleck90,Katsumata89,Chiba91}. 
However, the effect of a {\it staggered} field, that couples 
directly to the order parameter of the
classical system, has not been investigated in sufficient detail. This
is mainly due to the fact that a strong magnetic field modulated on the
microscopic scale was thought to be all but impossible to realize
experimentally \cite{comment1}. 
A breakthrough came with neutron scattering experiments
on $R_2$BaNiO$_5$ ($R$ = magnetic rare-earth) linear-chain nickelates
and their interpretation in terms of non-interacting Haldane spin chains
immersed in a strong effective staggered exchange field 
\cite{Zheludev96:PBANO-B,Maslov98}. In
$R_2$BaNiO$_5$ compounds almost perfectly isotropic antiferromagnetic
$S=1$ chains are formed by the Ni$^{2+}$ ions. The effective staggered
field is generated by the $R^{3+}$ sublattice that becomes ordered
magnetically below some Neel temperature $T_{N}$. The staggered field
intensity is proportional to the magnitude of the ordered moment of 
$R^{3+}$ magnetic ions and can be controlled in an experiment 
indirectly, by varying
the temperature. One of the most significant results 
was the first direct measurement of
the  staggered magnetization curve $M_s(H_s)$ of a Haldane spin
system\cite{Zheludev98:NBANO-L}.
It was found that of particular value as model systems
are (Nd$_x$Y$_{1-x}$)$_2$BaNiO$_5$ species, where the effective
interaction between $R$ and Ni-sublattices is of Ising
type\cite{Yokoo98}, 
so that the transverse excitations on the Haldane chains are effectively
decoupled from those on the $R$- subsystem. Thus the effect of the 
staggered field on transversal 
magnetic excitation spectrum of an 
isolated Ni$^{2+}$ Haldane chain could be measured 
experimentally\cite{Yokoo98}. 

In \cite{Maslov98} we gave a qualitative theoretical description of
quantum disordered antiferromagnets in the presence of a staggered
magnetic field and discussed the results in relevance to existing data
on $R_2$BaNiO$_5$ materials. The principal conclusion was that in a weak
staggered field the energy gap $\Delta$ increases in proportion to the
square of induced staggered moment on the Haldane chains. It was also
shown that a staggered field partially lifts the degeneracy of the
magnon triplet, the gap in the longitudinal mode being three times more
sensitive to $H_{s}$ than that in two transversal magnons. In the
present paper we refine this approach and obtain {\it quantitative}
predictions that we directly compare to recent experimental data for
(Nd$_x$ Y$_{1-x}$)$_2$BaNiO$_5$. We demonstrate that the
behavior of both transversal and longitudinal energy gaps in the
presence of a staggered field is contained in the staggered
magnetization curve $M_s(H_s)$. Our central result is that for a variety
of gapped one-dimensional antiferromagnets $M_s(H_s)$ has a {\it
universal} shape defined by three experimentally accessible parameters:
zero-field magnon energy gap $\Delta$, spin-wave velocity $v$, and the
renormalization constant $Z$, related to the residue of the magnon pole
at $H_s=0$.

A traditional theoretical description of spin dynamics
in one-dimensional quantum antiferromagnets is based on the mapping of these
systems to the 1+1-dimensional $O(3)$ Non-Linear Sigma Model (NLSM) 
with one spatial and one temporal coordinates.
This approach was
first introduced by Haldane \cite{Haldane83} for a Heisenberg
antiferromagnetic spin chain, and later extended to a variety of other
systems (for a recent review see \cite{Sierra96}). Although this
mapping is based on a quasiclassical approximation, and in principle it
should work well only for large spins $S \gg 1$, it gives
quantitatively correct predictions for any spin value.
In the absence of topological term
the NLSM Lagrangian in the presence of external (staggered)
field can be written as
\begin{equation}
{\cal L}= {1 \over 2g_0} \int dx
\left[{1 \over v_0} \left( {\partial \vec{\varphi}
\over \partial t} \right)^2- v_0 \left( {\partial \vec{\varphi}
\over \partial x} \right)^2 + 2 g_0 S \vec{H} _s \vec{\varphi} \right].
\label{nlsm}
\end{equation}
Here $\vec{\varphi}$ is a three component unit vector
($\vec{\varphi}^2=1$), pointing in the
direction of the local staggered moment,
$v_0$ is the bare spin wave velocity
of the system, and $g_0$ is the dimensionless
parameter, controlling the strength of
quantum fluctuations.
Notice that in our notation the staggered field $H_s$ is
coupled to the local staggered {\it spin}.
The staggered {\it magnetic} field $H_s^{(m)}$, coupled to
local staggered magnetic moment,  is proportional to $H_s$:
$g_m \mu_B H_s^{(m)}=H_s$, where $g_m$ is the
$g$-factor of a magnetic ion, and $\mu_B$ is the Bohr magneton.
In a large-S mapping of the Heisenberg spin chain
to the NLSM \cite{Haldane83} one has $v_0=2JS$, and $g_0=2/S$.
In a more general situation these parameters should be treated as
phenomenological constants fine-tuned to give the correct
low energy properties of the system.
The 1+1-dimensional $O(3)$  NLSM is always in a disordered state.
The correlation length in the absence of staggered field is
given by $\xi \sim a \exp(2 \pi/g_0)$, where
$a$ is the lattice spacing. The gap $\Delta$ in the
excitation spectrum is related to the correlation
length via a usual relation $\Delta = \hbar v/\xi$.

In order to calculate the macroscopic properties
of the system such as its staggered magnetization curve
or the excitation spectrum near the AFM zone center, one needs to
coarse-grain the NLSM Lagrangian by integrating out the large
$q$ and $\omega$ (small $x$ and $t$) degrees of freedom.
The change of parameters of the Lagrangian as a result
of this coarse-graining is described by the
Renormalization Group (RG) flow equations.
In real space the renormalization procedure
corresponds to replacing the
field $\vec{\varphi}(x,t)$ with $\vec{\varphi}_r(x,t)=(v/l^2)
\int_{|x'-x|<l, |t'-t|<l/v} \vec{\varphi}(x',t') dx'dt'$.
It is easy to see that the coarse-grained field
variable $\vec{\varphi}_r$ no longer
has a well defined length. In other words,
as a result of coarse-graining the
NLSM Lagrangian, characterized by a rigid constraint
$\vec{\varphi}^2=1$, becomes a ``soft-spin'' Lagrangian.
The length of coarse-grained field variable $\vec{\varphi}_r$
has a probability distribution, defined by some
effective potential.
The fully renormalized Lagrangian can be written as
\beq
{\cal L}= \int dx \left[{1 \over 2v} \left( {\partial \vec{\phi}
\over \partial t} \right)^2-{v \over 2} \left( {\partial \vec{\phi}
\over \partial x} \right)^2 - U(|\vec{\phi}|)+
\sqrt{Z}\vec{H_s}\vec{\phi} \right]
\label{lagr}
\eeq
In this expression we have introduced the {\it renormalized }
staggered field variable $\vec{\phi}$, defined through
$\sqrt{Z}\vec{\phi}= S\vec{\varphi}$.
The renormalization parameter $Z$ was
fine-tuned to give the desired form of the
derivative terms, with $v$ being the true 
spin wave velocity.
To specify this fully renormalized Lagrangian we need to know
the form of effective potential $U(|\vec{\phi}|)$.
For small $\vec{\phi}$ it
can be expanded in Taylor series
in  $\vec{\phi}^2$.
The quadratic term can be written as 
$(\Delta^2/2v)\vec{\phi}^2$, where $\Delta$ is the true
energy gap of the magnon spectrum. Indeed, the 
quadratic part of \req{lagr} should describe a triplet of
``relativistic'' non-interacting bosons. The
spin wave velocity $v$ plays the role of the speed of
light in its relativistic analogue.
The above form of the quadratic term correctly reproduces
the spin-wave dispersion
${\cal E}(\pi+k)=\sqrt{\Delta^2+v^2 k^2}$.
The relativistic analogy also determines the intensity
of the magnon pole in the spin correlator as
\beq
S^{\alpha \beta}_{SMA}(q, \omega)=
\delta_{\alpha \beta}{Zv \over 2{\cal E}(q)} 2 \pi \delta(\omega-{\cal
E}(q)).
\label{pole}
\eeq
In 1+1 dimensions
it is convenient to write the Taylor expansion
of $U(|\vec{\phi}|)$ in terms of the set of  dimensionless
parameters $u_{2n}$ defined by
\beq
U(|\vec{\phi}|)=
{\Delta ^2 \over v} \big(
{1 \over 2} |\vec{\phi}|^2
+{1 \over 4!} u_4 |\vec{\phi}|^4
+{1 \over 6!} u_6 |\vec{\phi}|^6
%+
%{1 \over 8!} u_8 |\vec{\phi}|^8 \big)
%+o(|\vec{\phi}|^{10})
+\ldots \big)
\eeq
The effective potential, truncated at the $|\vec{\phi}|^4$
term,  corresponds to the Ginzburg-Landau Lagrangian
introduced by Affleck \cite{Affleck89}
on phenomenological grounds to describe Haldane-gap systems.
%However, to the best of our knowledge, the strength $\lambda$
%of the {\it positive} quartic term
%$\lambda |\vec{\phi}|^4$ was never calculated in the literature
%on this subject. The analysis of the higher order terms
%of the Taylor expansion of the effective potential was
%also never attempted in this context.

The {\it linear} staggered susceptibility at zero external staggered field
can be derived from the quadratic part of Lagrangian 
describing non-interacting magnons. 
Indeed, the expectation value of $\vec{\phi}$ in the
presence of a weak staggered field is easily obtained by
balancing the quadratic term and the source term in Eq.\req{lagr}.
The result is the single mode contribution
to zero-field staggered susceptibility
\beq
\chi^{(s)}(0)={Zv \over \Delta^2} \qquad .
\eeq
We can compare this prediction to the numerical results
for the S=1 Heisenberg chain.
Using the numerical values $Z=1.26$ ($g$ in their
notation), $v=2.49J$, and $\Delta=0.41J$, reported in \cite{Sorensen94},
the single mode approximation gives $\chi^{(s)}(0) \simeq  18.7/J$
in excellent agreement with
Monte Carlo result $21(1)/J$ \cite{Sakai94}.
This is a manifestation of the well known fact that
in Haldane-gap systems virtually all spectral weight at
the AFM zone center is concentrated in the
magnon triplet.

The main concern of the present paper is the {\it nonlinear} behavior in
arbitrary strong staggered fields. In order to describe these effects
quantitatively one needs to know the numerical values of dimensionless
couplings $u_{2n}$, which in principle should depend on the parameters
of the bare Lagrangian. If, however, the correlation length is
sufficiently long one can safely assume that these couplings are at
their RG fixed point values. Fortunately, very accurate numerical values
of the universal fixed point couplings $u_4$, $u_6$, and $u_8$ were
recently obtained by Pelissetto and Vicari in
\cite{Pelissetto97,Pelissetto98}. They carefully compared the results of
$\epsilon$, $1/N$, high temperature, and strong coupling expansions of
$O(N)$-symmetric models in $d$ dimensions with the results of Monte
Carlo simulations. For $d=2$, $N=3$ their estimates are $u_4=11.8(1),
u_6=3.33(10) \times u_4^2=460(20),u_8=20(5) \times u_4^3=33,000(8,000)$
\cite{comment-Maslov98:L}. In  \cite{Pelissetto97} the deviations of
couplings from their fixed point values were estimated as
$u_{2n}(\xi)-u_{2n}(\infty) \sim 1/\xi^2$.
For the $S=1$ Heisenberg antiferromagnet the correlation length
$\xi \simeq 6$ (in units of lattice spacing). Therefore,
one could reasonably expect the deviation of $u_{2n}$
from their fixed point values to be around $1/36 \simeq 3\%$!
This fact is confirmed numerically in Fig. 2 of \cite{Pelissetto97}.

From the above it follows that the quartic term 
$\lambda |\vec{\phi}|^4$ used by Affleck \cite{Affleck89} 
to describe the effects of pairwise magnon repulsion has in fact 
a universal strength $\lambda=(u_4/4!) (\Delta^2/v) \simeq 0.49 \Delta^2/v$. 
In their theoretical study of the effect of the external 
field on the excitation spectrum in NENP Mitra and Halperin
\cite{Mitra94} made a rough estimate of the value of $\lambda$. By
matching the first term in perturbative $1/H_s$ high field expansion of
the magnetization function with a small $H_s$ expansion they have got an
order of magnitude estimate, which in our notation corresponds to
$\lambda \simeq Z \Delta^2/4v \sim 0.31
\Delta^2/v$ rather close to our more refined result.

The effective potential manifests itself in the
staggered magnetization curve of a 1D gapped antiferromagnet.
Let us select $z$-axis along the external staggered field.
The expectation value of the field $\vec{\phi}$
is defined by the minimum of the total potential energy
(effective potential plus an external field term) located at
$\sqrt{Z} H_s=U'(\avg{\phi_z})$.
Therefore, the staggered
magnetization curve $M_s(H_s)$ is defined by the equation
\beqar
H_s&=&{1 \over \sqrt{Z}} U'\left({M_s \over \sqrt{Z}}
\right)= \nonumber \\
&=&M_s {\Delta^2 \over Z v} \left[ 1+{u_4 \over 3!}{M_s^2 \over Z}+
{u_6 \over 5!}{M_s^4 \over Z^2}+ {u_8 \over 7!}
{M_s^6 \over Z^3}+\ldots \right] .
\label{magn_curve}
\eeqar
Using the numerical estimates for $u_4$, $u_6$, $u_8$ quoted above, and
$Z=1.26$ from \cite{Sorensen94}, for $S=1$ chain we obtain
$\chi^{(s)}(0) H_s=M_s(1+1.56 M_s^2 + 2.4 M_s^4 + 3.27 M_s^6)$ This
result can be directly compared to the $M_{s}(H_{s})$ curve measured
experimentally in Nd$_x$Y$_{1-x}$)$_2$BaNiO$_5$ \cite{Zheludev98:NBANO-L} 
(Fig. 1). The solid line is a fit to the experimental data with
$\chi^{(s)}(0)$ being the only adjustable parameter. We note that
the result of the fit -- $\chi^{(s)}_{\rm exp}(0)=0.53$meV$^{-1}$ 
differs from the expected value 
$\chi^{(s)}_{\rm theor}(0)=18.7/J \simeq 0.85$meV$^{-1}$. 
It has to be emphasized
however, that the experimental scaling of the abscissa in Fig. 1 in units of
magnetic field heavily relies on a series of assumptions and
simplifications regarding the properties of rare earth ions, and the
exact applicability of the mean-field model in a wide temperature
range \cite{Zheludev98:NBANO-L}.  The experimentally determined {\it shape} of
$M_{s}(H_{s})$, on the other hand, is a more robust result and is in
excellent agreement with our theoretical predictions.

The effective potential also contains information on the behavior of the
excitation spectrum in the presence of a finite staggered field. The
magnon excitations are by definition deviations of local staggered
magnetization from its equilibrium value. When the $O(3)$ symmetry is
broken by the external field, the degeneracy of the magnon triplet is
partially lifted. In this case one should distinguish between the gap
$\Delta_{||}$ in the longitudinal branch of $\delta \phi _z$ excitations
and the transversal gap $\Delta_{\perp}$ in the doublet of  $\delta \phi
_x$, $\delta \phi _y$ excitations.
Both these gaps are determined by the quadratic terms in the expansion of 
$%$
U(|\vec{\phi}+\delta \vec{\phi}|) \simeq U(|\vec{\phi}|)+
\sum (\Delta_{\alpha}^2/2v)\ \delta \phi_{\alpha}^2 
$.  %$
After some straightforward algebra one gets
$%$
U(|\vec{\phi}+\delta \vec{\phi}|)=
U \big(\sqrt{(\avg{\phi_z}+\delta \phi_z)^2+ \delta
\phi _x^2+\delta \phi _y^2}\big) \simeq U(\avg{\phi_z})+
U'(\avg{\phi_z}) \delta \phi _z + (U'(\avg{\phi_z})/2\avg{\phi_z})
(\delta \phi _x^2+\delta \phi _y^2)+(U''(\avg{\phi_z})/2)\delta
\phi_z^2
$. %$
The term linear in $\delta \phi _z$ is precisely
compensated by $-\sqrt{Z}H_s \delta \phi _z$ coming from
the external field term.
The above expression for the quadratic terms in the expansion of 
$U(|\vec{\phi}+\delta \vec{\phi}|)$ should not come as 
a big surprise. Indeed, for the isotropic system
the transversal staggered susceptibility in the presence of a finite
staggered field $H_s$ is given by $\chi_{\perp}^{(s)}=M_s(H_s)/H_s$,
while the longitudinal staggered susceptibility is
$\chi_{||}^{(s)}=dM_s(H_s)/dH_s$.
On the other hand,
in the single mode approximation one has 
$\chi_{\alpha}^{(s)} (H_s)=Z v /\Delta_{\alpha}^2(H_s)$.  
This argument again fixes the 
longitudinal and transversal energy gaps at  
\beqar
\Delta_{\perp}^2(M_s)&=&v{\sqrt{Z} U'(M_s/\sqrt{Z}) \over
M_s}= \nonumber \\
&=&
\Delta ^2 \left[ 1+{u_4 \over 3!}{M_s^2 \over Z}+
{u_6 \over 5!}{M_s^4 \over Z^2}+\ldots \right];
\label{d_perp_1} \\
\Delta_{||}^2(M_s)&=&v U''(M_s/\sqrt{Z})= \nonumber \\
&=&
\Delta ^2 \left[ 1+{u_4 \over 2!}{M_s^2 \over Z}+
{u_6 \over 4!}{M_s^4 \over Z^2}+\ldots \right];
\label{d_par_1}
\eeqar

At this point it is important to see what assumptions the derivation
of Eqs.(\ref{d_perp_1}, \ref{d_par_1}) relies on. Indeed, we
implicitly assumed that the only parameter of the system which changes
in the presence of staggered field are energy gaps $\Delta_{\perp}$ ,
and $\Delta_{||}$. We have disregarded the changes in $Z$ and $v$ with
field. It is well known that the spin wave velocity is not very
sensitive to the parameters of the system and changes
only slightly when, for instance, anisotropy is switched on
\cite{Sorensen94}. This is rather natural result in field
theoretical formulation of the problem since the velocity should not
be renormalized at all except due to intrinsic microscopic asymmetry of
spatial and temporal directions (spatial coordinate has underlying
discrete lattice structure, while time is naturally continuous).
The weakness of the change of $Z$ with the field is due to the fact that
in two dimensions the critical exponent $\eta$, relating $Z$ to
the correlation length $\xi$ as $Z \sim \xi^{-\eta}$,
is equal to zero. Connecting $Z$ to its
NLSM counterpart $g_0$ in \req{nlsm}
one can approximately write $\xi \sim \exp (-2
\pi /Z)$ or, alternatively, $Z^{-1}(\xi)={\rm const}-(2 \pi)^{-1} \ln
\xi={\rm const}+(2 \pi)^{-1} \ln \Delta $. We see that a change in the
gap $\Delta$  leads only to the logarithmic change in $Z$. Such
corrections are disregarded at the precision of our calculations.
In \req{lagr} the weak dependence of $Z$ on $\xi$ is reflected in the 
omission of the nonlinear terms
containing derivatives of $\vec{\phi}$. 
It is a well know fact \cite{Berges96} that for small values of
$\eta$ this is a valid approximation.
We conclude with the derivation of an explicit expression for the
staggered field dependence of longitudinal and transversal gaps for
$S=1$ Heisenberg chain. Plugging the numerical values for $g_4$,
$g_6$, $g_8$, and $Z$ in Eqs. (\ref{d_perp_1}, \ref{d_par_1}) we
get
\beqar
{\Delta_{\perp}^2 (M_s) \over \Delta ^2}
&=& 1+1.56 M_s^2 + 2.4 M_s^4 + 3.27 M_s^6
\label{d_perp_2} \\
{\Delta_{||}^2 (M_s) \over \Delta ^2 }
&=& 1+4.68 M_s^2 + 12 M_s^4 + 22.9 M_s^6
\label{d_par_2}
\eeqar
As shown in Fig. 2, our prediction for the transversal gap is in
excellent agreement with neutron scattering data on (Nd$_x$
Y$_{1-x}$)$_2$BaNiO$_5$ family of compounds \cite{Yokoo98},  
with no adjustable parameters.

The work at Brookhaven National Laboratory was
supported by the U.S. Department of Energy Division
of Material Science, under contract DE-AC02-98CH10886.

%\bibliographystyle{prsty}
%\bibliography{/export/home/cmt3.phy/maslov/BIB/haldane}

\begin{thebibliography}{10}
\bibitem{comment1}
The effects of a weak staggered component mixed to 
a strong uniform field were experimentally 
observed in [5].
\end{thebibliography}

\begin{figure}
\caption{The staggered magnetization curve as deduced from neutron powder
diffraction data \protect\cite{Zheludev98:NBANO-L}) 
on (Nd$_x$Y$_{1-x}$)$_2$BaNiO$_5$ ($x=0.25,0.5,1$).
The effective staggered field is estimated through 
a mean-field analysis \protect\cite{Zheludev98:NBANO-L}.
The solid line is a single-parameter fit with our theoretical result
$\chi^{(s)}(0) H_s=M_s(1+1.56 M_s^2 + 2.4 M_s^4 + 3.27 M_s^6)$~ (see text).}
\end{figure}

\begin{figure}
\caption{Inelastic neutron scattering data for the relative
increase in the energy gap
of the Ni-chain magnon as a function of induced staggered spin
$M_s$ in Nd$_x$Y$_{1-x}$)$_2$BaNiO$_5$ 
\protect\cite{Zheludev98:NBANO-L}. 
Solid lines are defined by (9,10)
The experimental data agree with the theoretical
prediction for the {\it transversal} gap.}
\end{figure}

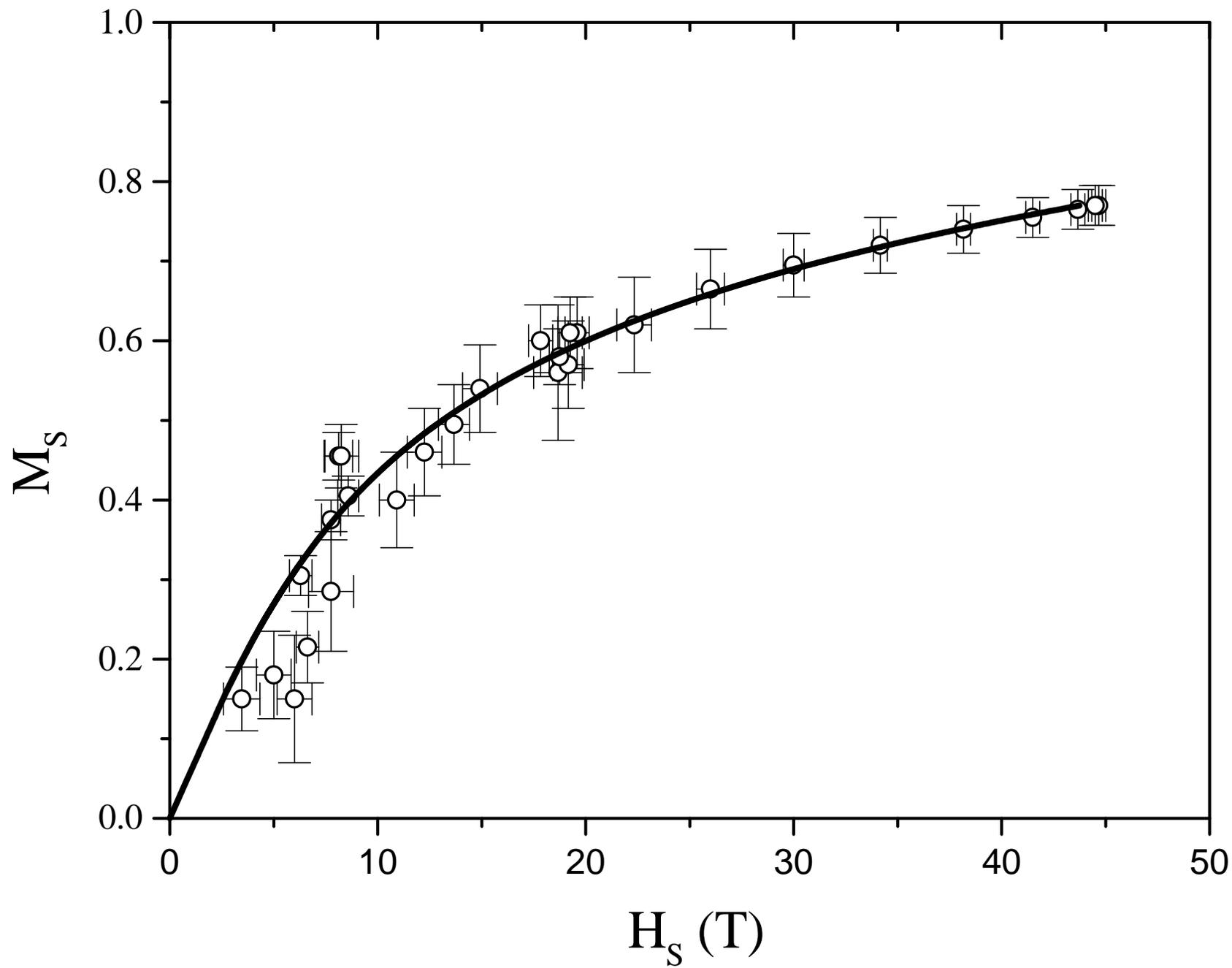
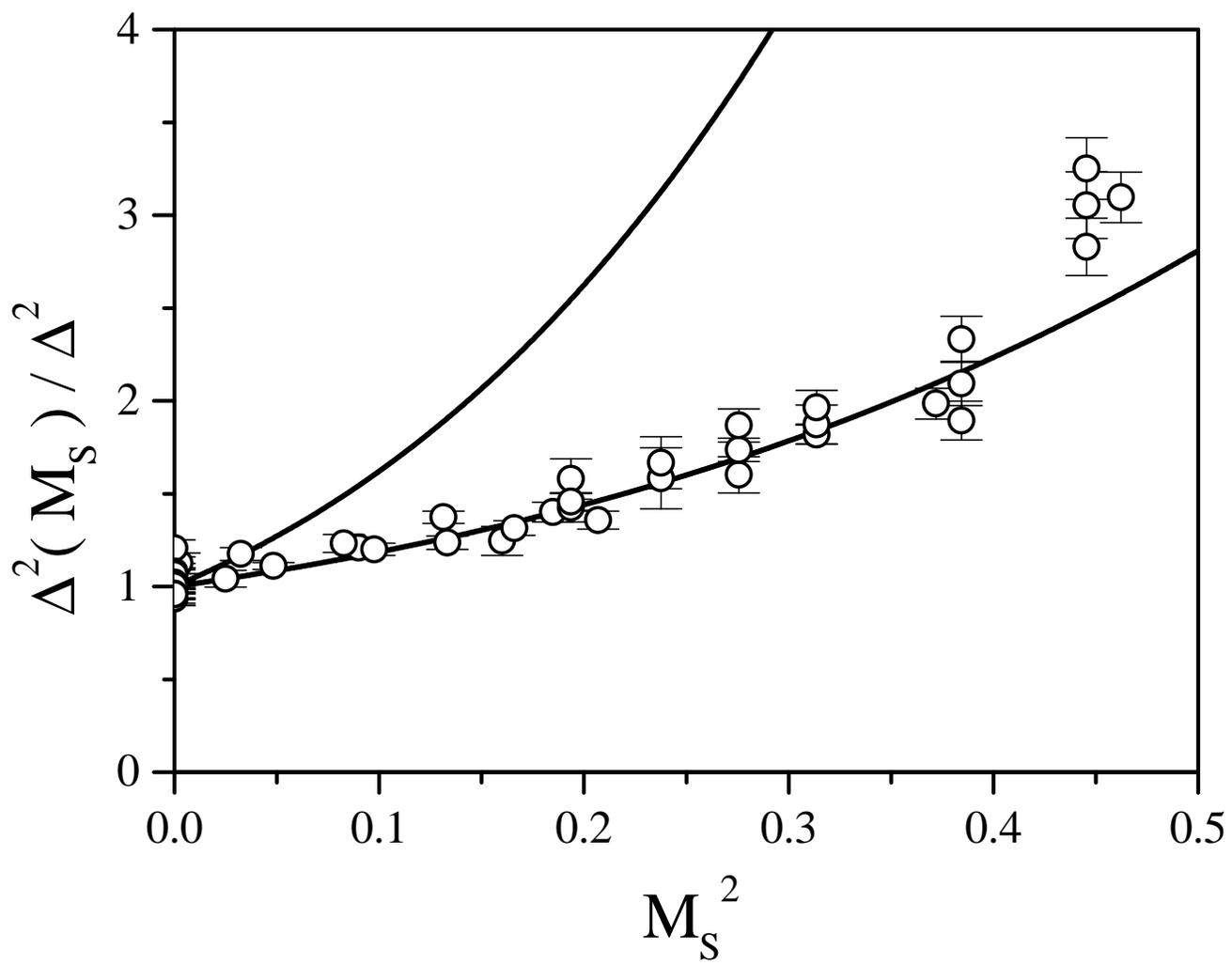
\end{document}